\documentstyle[12pt,aps,epsf]{revtex}
\def\lQ{\Lambda_{\rm QCD}}

\def\als{\alpha_{\rm s}}   
\def\siml{{\ \lower-1.2pt\vbox{\hbox{\rlap{$<$}\lower6pt\vbox{\hbox{$\sim$}}}}}}

\begin{document}
\tolerance=10000
\hfuzz=5 pt
\baselineskip=24 pt
\draft
\preprint{CERN-TH/2000-036}
\title{The $B_c$ mass up to order $\alpha_{\rm s}^4$}
\vskip 0.25cm
\author{Nora Brambilla  and Antonio  Vairo}
\address{Theory Division CERN, 1211 Geneva 23, Switzerland}
\maketitle
\vspace{0.5 cm}
\begin{abstract}
\noindent
We evaluate in perturbative QCD, up to order $\alpha_{\rm s}^4$, the mass of the $B_c$. 
We use the so-called $1S$-mass in order to improve the convergence of the perturbative series. 
Our result is $E(B_c)_{\rm  pert} = 6326^{+29}_{-9} \, {\rm MeV}$. 
Non-perturbative effects are discussed. A comparison with potential models
seems to be consistent with non-perturbative contributions of the order 
$ - (40\div 100) \, {\rm MeV}$. \vspace{2mm} 
\end{abstract}

\section{Introduction}
The discovery of the $B_c$ meson (the lowest pseudoscalar $\,^1S_0$ state of the 
$\bar{b}c$ system) has been reported in 1998 by the CDF collaboration in the 1.8 TeV 
$p \bar{p}$ collisions at the Fermilab Tevatron \cite{exp}. The  mass has been measured to be 
$6.40\pm 0.39\pm 0.13$ GeV. 

The fact that the mass of the quarks of quark--antiquark systems 
built up by $b$ and $c$ quarks is much larger than the typical binding energy 
suggests that these systems are non-relativistic, i.e. that the 
heavy-quark velocity $v$ is small.  The typical scales of these systems are the binding 
energy $\sim m v^2$ and the momentum transfer $\sim m v$; moreover, 
because of the non-relativistic nature of the system,  $m \gg mv \gg mv^2$
(for the purpouses of this discussion $m$ and $v$ can be identified with the
mass and the velocity of the lightest component of the bound state respectively). 
Let us call $\lQ$ the scale at which non-perturbative effects become important.

If $\lQ \siml \, m v^2$, then the scale $m v$ can be integrated out order 
by order in $\alpha_{\rm s}$ at a scale $mv \gg \mu^\prime \gg mv^2$. 
The system is described up to order $\alpha_{\rm s}^4$ by a potential which 
is entirely accessible to perturbative QCD and at the leading order is the Coulomb potential. 
Non-potential effects start at order $\alpha_{\rm s}^5\ln \mu^\prime$ \cite{pnrqcd2,pnrqcd1}. This kind of system 
is called Coulombic. Non-perturbative effects are of non-potential type.
In the particular situation $m v^2 \gg \lQ$ they can be encoded into local condensates \cite{Voloshin}. 
This condition seems to be fulfilled by the bottomonium ground state, which has been 
studied in this way in \cite{Pineda}. Also the charmonium ground state has been analysed 
as a Coulombic bound state by the same authors. In both cases 
(but with caveats) the non-perturbative corrections 
\`a la Voloshin--Leutwyler, i.e. in terms of local condensates, 
have been claimed to be under control \cite{Pineda2}. 

For heavy quarkonium states higher than the ground state the condition $\lQ \siml \, m v^2$ 
is not fulfilled and non-perturbative terms affect the potential. The system is no longer 
Coulombic. Traditionally the energy of these systems has been calculated within QCD-inspired  
confining potential models. A large variety of them exists in the literature 
and they have been on the whole quite successful (cf. \cite{pot} for some recent reviews).
However, the usual criticisms apply. Their connection with the QCD parameters is hidden, 
the scale at which they are defined is not clear, they cannot be systematically improved
and they usually contain a superposition by hand of perturbative and non-perturbative effects. 
For this reason a lot of effort has been devoted, over the years, to obtaining the relevant 
potentials from QCD  by relating them to some Wilson loops expectation values \cite{wilson,pot}. 
Anyway, these have to be eventually computed either via lattice simulations 
or in QCD vacuum models \cite{Brambilla}. In the specific situation $mv \gg \lQ \gg  m v^2$ 
the scale $mv$ can still be integrated out perturbatively, giving rise to a Coulomb-type potential. 
Non-perturbative contributions to the potential will arise when integrating out the scale $\lQ$. 
This situation has been studied in \cite{pnrqcd2}. We will call quasi-Coulombic the systems described 
by the situation $mv \gg \lQ \gg m v^2$, when the non-perturbative piece of the potential 
can be considered small with respect to the Coulombic one and treated as a perturbation.

The only available theoretical predictions (to our knowledge) of the $B_c$ mass resort to  
(confining) potential models or to the lattice. In this work we will carry out
the calculation of the perturbative $B_c$ mass up to order $\alpha_{\rm s}^4$. 
We will call it $E(B_c)_{\rm  pert}$\footnote{It is in general somehow ambiguous to 
separate in a physical quantity perturbative from non-perturbative contributions. 
From this point of view the following Eq. (\ref{EBc}) may be seen as a 
definition of what we call here perturbative $B_c$ mass. Analogous definitions have to be 
understood for the perturbative $J/\psi$ and $\Upsilon(1S)$ masses.}. This calculation 
will be relevant to a QCD determination of the $\bar{b}c$ ground state if this system is Coulombic 
or at least quasi-Coulombic. Moreover, in the way we are doing the calculation, we also assume 
the $\Upsilon(1S)$ and the $J/\psi$ to be Coulombic or at least quasi-Coulombic systems.
The question if these assumptions correspond to the actual systems cannot be settled 
at this point. On the other hand there is no {\it a priori} reason to rule them out. 
Let us consider, for instance, the argument used in \cite{Pineda} for the $J/\psi$.
Lattice data show that the static potential clearly deviates from a $1/r$
behaviour for distances larger than 1 GeV$^{-1}$ (see \cite{pot}
and references therein). Therefore the $J/\psi$ is Coulombic or quasi-Coulombic 
if the characteristic scale of the bound state, $\mu \sim m_c v_c$, is bigger than 1 GeV. 
If we assume $m_c \simeq (1.6 \div 2.0)$ GeV and if we fix that scale on the Bohr radius, 
$a$, of the $J/\psi$, $\mu = 2/a(\mu)$, then we get $\mu \simeq (1.5 \div 1.6)$ GeV. 
Since this scale falls into the energy window between the mass scale $m_c$ and 1 GeV, these
figures are consistent with a Coulombic or quasi-Coulombic picture of the $J/\psi$ \footnote{
One may wonder if these figures are consistent with the non-relativistic expansion 
underlying NRQCD. Only an actual calculation may decide this, 
since a break-down of the NRQCD expansion, if it occurs, should be manifest 
in a breakdown of the expansion of the energy levels. In the specific situation of the $B_c$, 
as we will see later on in this paper, the expansion that we get shows a still convergent behaviour. }. 

The main problem of the calculation of the perturbative $B_c$ mass 
is the well-known bad convergence of the perturbative series when using the
pole mass. This is due to a renormalon cancellation 
occurring between the pole mass and the static Coulomb potential \cite{beneke}.
We handle the problem by expressing the $c$ and the $b$ pole mass in the perturbative 
expression of the $B_c$ mass as half of the perturbative mass of the $J/\psi$ 
($E(J/\psi)_{\rm  pert}$) and the $\Upsilon(1S)$ ($E(\Upsilon(1S))_{\rm  pert}$) respectively.  
This corresponds to using the quark mass in the so-called $1S$ scheme introduced in \cite{manohar}.
In this way, by expressing $E(B_c)_{\rm  pert}$ in terms of quantities that are infrared safe at order 
$\Lambda_{\rm QCD}$ (the $^3S_1$ perturbative masses), the pathologies of the perturbative series, 
due to the renormalon ambiguities affecting the pole mass, are cured. We will explicitly show that, in fact,   
we obtain a better convergence of the perturbative expansion and   
a stable determination of the perturbative mass of the $B_c$, just in the 
energy range that the above discussion on the
$J/\psi$ suggests to be also the relevant one for the $B_c$. 
Non-perturbative terms are of potential type in the quasi-Coulombic situation and 
of non-potential type in the Coulombic situation. They affect the identification of 
the perturbative masses $E(J/\psi)_{\rm  pert}$, $E(\Upsilon(1S))_{\rm  pert}$ 
and $E(B_c)_{\rm  pert}$ with the corresponding physical ones. 
If we aim at obtaining a good estimate of the physical $B_c$ mass, 
it is not important for each of these contributions to be individually small, as long as 
the sum of them in the $B_c$ mass is small. As we will discuss at the end, a picture with 
non-perturbative corrections to the $B_c$ mass of a not too large size ($\siml\, 100$ MeV) 
seems to be consistent with the experimental data and with the potential models. 

The paper is organized in the following way. In the next section we set up the 
formalism and perform the calculation of the perturbative $B_c$ mass. 
In section 3 we briefly discuss the non-perturbative corrections and compare our result with  other 
determinations of the $B_c$ mass available in the literature.

\section{Calculation of $E(B_c)_{\rm  pert}$}
In order to calculate the $B_c$ mass in perturbation theory up to order $\alpha_{\rm s}^4$, 
we need to consider the following contributions to the potential: the perturbative 
static potential at two loops, the $1/m$ relativistic corrections at one loop,   the spin-independent 
$1/m^2$ relativistic corrections at tree level and the $1/m^3$ correction to the kinetic energy. 
We will not consider here effects due to a non-zero charm quark mass on the $b$ (and the $b\bar{b}$ system) 
of the type discussed in \cite{cloop}. 
We will follow the derivation of the heavy quarkonium mass of Ref. \cite{Pineda}. 

The static potential at two-loops has been calculated in \cite{Peter}.
It is useful, in order to perform an analytic calculation, to split it as 
$$
V_0(r) = v_0(r)  + \delta v_0(r), 
$$
where $v_0$ is the part that does not contain logarithms, 
\begin{eqnarray*}
& & \hspace{-8mm}
v_0(r) \equiv -C_F {\tilde\alpha_{\rm s}(\mu)\over r},\\
& & \hspace{-8mm} \tilde\alpha_{\rm s}(\mu) \equiv \alpha_{\rm s}(\mu) 
\left\{1 + {\alpha_{\rm s}(\mu) \over \pi}\left[ {5 \over 12} \beta_0  - {2\over 3}C_A + 
{\beta_0 \over 2} \gamma_E \right] \right.
\nonumber\\
& & \hspace{2mm}
\left. + \left({\alpha_{\rm s} \over \pi}\right)^2 \left[ \beta_0^2 \left( {\gamma_E^2\over 4} 
+ {\pi^2\over 48}\right) + \left( {\beta_1\over 8} + {5\over 12} \beta_0^2 - {2\over 3} C_A \beta_0 \right)
\gamma_E + {c\over 16} \right]\right\},
\nonumber
\end{eqnarray*}
where $\beta_n$ are the $\beta$-function coefficients\footnote{
$\beta_0 = 11 C_A/3 - 2/3 N_f$, $\beta_1 = 34 C^2_A/3 - 10 N_f C_A/3 -2 N_f C_F$, ... }, 
$c \equiv  \displaystyle\left({4343\over 162} + 4\pi^2 - {\pi^4\over 4} + {22\over 3}\zeta_3 \right)C_A^2$ 
$-$ $\displaystyle\left( {899\over 81} + {28\over 3}\zeta_3 \right) C_A N_f$ 
$-$ $\displaystyle\left({55\over 6} - 8 \zeta_3 \right) C_F N_f$  
$+$ $\displaystyle\left({10\over 9} N_f \right)^2$, $\gamma_E = 0.5772\dots$ is the Euler constant,
$C_A=3$, $C_F=4/3$ and $N_f$ is the number of flavours (we will take $N_f =3$)\footnote{
The result one obtains by choosing $N_f = 4$ and 
$\Lambda_{\overline{\rm MS}}^{N_f=4}$ $=$ 230 MeV has been checked 
to be consistent with the central value and the errors of Eq. (\ref{ebc}).}; $\delta v_0$ is given by  
\begin{eqnarray*}
\delta v_0(r) \equiv  - {C_F\alpha_{\rm s}(\mu)^2\over \pi r} \ln (\mu r) 
\left\{  {\beta_0 \over 2} + {\alpha_{\rm s} \over \pi} \left[ \beta_0^2 {\ln(\mu r) + 2 \gamma_E \over 4} 
+ {\beta_1\over 8} \right.\right. 
\left.\left.+ {5\over 12} \beta_0^2 - {2\over 3} C_A \beta_0 \right]\right\}.
\end{eqnarray*}
The strong coupling constant $\alpha_{\rm s}$ is understood in the $\overline{\rm MS}$ scheme. 
At the scale $\mu$ we will take the value of $\alpha_{\rm s}$ from the three-loop expression with 
$\Lambda_{\overline{\rm MS}}^{N_f=3}$ $=$ $(300 \pm 50)$ MeV. The $1/m$ relativistic corrections 
at one loop,  the $1/m^2$ tree-level spin-independent terms and the $1/m^3$ correction 
to the kinetic energy are given by \cite{Gupta}
\begin{eqnarray*}
\delta v_1(r) &\equiv& 
\left( 2C_F^2 {m_{\rm red}\over  m_b m_c}  - {C_A C_F\over m_{\rm red}} \right) {\alpha_{\rm s}^2\over 4 r^2} 
+ {C_F\alpha_{\rm s}\over m_b m_c}{1\over r} \Delta \\
& & + {1\over 2}\left( {1\over m_b^2} + {1\over m_c^2} - {2 \over m_b m_c} \right) C_F \pi \alpha_{\rm s} 
\delta^{(3)}({\bf r}) - {\Delta^2\over 8}\left({1\over m_b^3} + {1\over m_c^3} \right). 
\end{eqnarray*}
$m_b$, $m_c$ and $m_{\rm red} \equiv m_bm_c/(m_b+m_c)$ 
are the $b$, the $c$ and the reduced pole mass respectively. 
Then, the Hamiltonian relevant in order to get the $B_c$ mass at $\alpha^4_{\rm s}$ accuracy is 
\begin{equation}
H(B_c) = m_b + m_c + {{\bf p}^2\over 2 m_{\rm red}} + v_0(r) + \delta v_0(r) + \delta v_1(r) .
\label{bc2}
\end{equation}

Up to order $\alpha^4_{\rm s}$ the ground-state energy is given by 
($\langle ~~ \rangle$ means the average on the ground state):  
\begin{equation}
\hspace{-8mm}
E(B_c)_{\rm  pert} \!=\! m_b + m_c - m_{\rm red} {(C_F \tilde\alpha_{\rm s}(\mu))^2\over 2} 
+ \langle \delta v_0 \rangle + \langle \delta v_1 \rangle 
+ \langle \delta v_0 G_c \delta v_0 \rangle, 
\label{EBc}
\end{equation}
where at leading order 
\begin{eqnarray*}
\langle \delta v_0 G_c \delta v_0 \rangle = - m_{\rm red}{C_F^2 \beta_0^2 \alpha_{\rm s}^4 \over 2\pi^2}
\left( {3 + 3 \gamma_E^2 -\pi^2 + 6 \zeta(3) \over 12} \right.
\left. - {\gamma_E\over 2} \ln (\mu a/2) + {1\over 4}\ln^2(\mu a/2) \right),
\end{eqnarray*}
with $a(\mu) \equiv 1/(m_{\rm red}C_F \tilde\alpha_{\rm s}(\mu))$, the Bohr radius of the system.
This corresponds to the only contribution  relevant at order $\alpha_{\rm s}^4$ produced by the Hamiltonian 
(\ref{bc2}) in second-order perturbation theory  ($G_c$ stands for the Coulombic 
intermediate states) and can be read off from the second reference in \cite{Pineda}.
The other averages can be easily evaluated by means of the standard formulas 
\begin{eqnarray*}
& &\hspace{-8mm}
\left\langle {1\over r} \ln^2(\mu r) \right\rangle = 
{1\over a} \left( \ln^2(\mu a/2) + 2(1-\gamma_E)\ln(\mu a/2) + (1-\gamma_E)^2 + {\pi^2\over 6} -1 \right),\\
& &\hspace{-8mm}
\left\langle {1\over r} \ln(\mu r) \right\rangle = 
{1\over a} \left( \ln(\mu a/2) -\gamma_E + 1\right),
\quad \left\langle {1\over r} \right\rangle = {1\over a}, 
\quad \left\langle {1\over r^2} \right\rangle = {2\over a^2}, 
\\
& &\hspace{-8mm} \left\langle {1\over r}\Delta  \right\rangle = - {3\over a^3}, 
\quad \langle \Delta^2 \rangle = {5\over a^4}, 
\quad \langle \delta^{(3)}({\bf r}) \rangle = {1\over \pi a^3}.
\end{eqnarray*}
After an  explicit calculation we get from Eq. (\ref{EBc}), up to order $\alpha_{\rm s}^4$: 
\begin{eqnarray}
& &\hspace{-10mm}
E(B_c)_{\rm  pert} \!=\! m_b \!+\! m_c \! + E_0(\mu)\left\{\! 1 \! - {\alpha_{\rm s}(\mu)\over \pi} \left[ \beta_0  
\ln\left({2 C_F \alpha_{\rm s} m_{\rm red}\over \mu}\right)      
+ {4\over 3} C_A - {11 \over 6} \beta_0 \right] \right.
\nonumber\\
& &\hspace{-10mm}
+ \left({\alpha_{\rm s}\over \pi}\right)^2 \left[ {3\over 4} 
\beta_0^2 \ln^2\left({2 C_F \alpha_{\rm s} m_{\rm red}\over \mu}\right)
+ \left(2 C_A \beta_0 - {9\over 4} \beta_0^2 - {\beta_1\over 4} \right) 
\ln\left({2 C_F \alpha_{\rm s} m_{\rm red}\over \mu}\right)  \right.
\nonumber\\
& &\hspace{-5mm}
- \pi^2 C_F^2  \left( {1\over m_b^2} + {1\over m_c^2} - {6\over m_b m_c} \right) m_{\rm red}^2  
+ {5\over 4} \pi^2 C_F^2  \left( {1\over m_b^3} + {1\over m_c^3}\right) m_{\rm red}^3    
\nonumber\\
& &\hspace{-5mm} 
\left.\left. + \pi^2 C_F C_A   + {4\over 9} C_A^2 - {17 \over 9} C_A \beta_0 
+ \left( {181 \over 144} + {1\over 2} \zeta(3) +{\pi^2\over 24} \right) \beta_0^2 
+ { \beta_1 \over 4} + {c\over 8} \right]\right\} ,
\label{EBc2}
\end{eqnarray}
where $E_0(\mu) = - m_{\rm red} \displaystyle{(C_F \alpha_{\rm s}(\mu))^2\over 2}$. 

The main problem connected with the perturbative series (\ref{EBc2}) is the  bad convergence 
in terms of the heavy-quark pole masses. Let us consider, for instance,
$\mu=1.6$ GeV, $m_b=5$ GeV and $m_c = 1.8$ GeV.  
Then we get $E(B_c)_{\rm  pert} \simeq 6149$ MeV $\simeq 6800 - 115  - 183 - 353$ MeV, 
where the second, third and fourth figures are the corrections of order $\alpha_{\rm s}^2$,  
$\alpha_{\rm s}^3$ and $\alpha_{\rm s}^4$ respectively. The series turns out to be very badly convergent. 
This reflects also in a strong dependence on the normalization scale $\mu$:
at $\mu = 1.2$ GeV we would get $E(B_c)_{\rm  pert} \simeq 5860$ MeV, while at 
$\mu = 2.0$ GeV we would get $E(B_c)_{\rm  pert} \simeq 6279$ MeV.\footnote{The result also 
depends on the $c$ and $b$ pole masses, which are poorly known. See the
following discussion.} 
The origin of this behaviour can be understood in the renormalon language. 
The pole mass is affected by an IR renormalon ambiguity that  
cancels against an IR renormalon ambiguity of order $\Lambda_{\rm QCD}$ present 
in the static potential \cite{beneke}. The non-convergence of the perturbative 
series (\ref{EBc2}) signals the fact that large $\beta_0$ contributions (coming 
from the static potential renormalon) are not summed up and cancelled against the pole masses.  
A possible solution, in order to avoid large perturbative 
corrections and large cancellations, or, in other words, in order to obtain a well-behaved 
perturbative expansion, is to resort to a different definition of the mass.
The so-called $1S$ mass of a heavy quark $Q$ is defined as half of the perturbative contribution 
of the $\, ^3S_1$ $Q-\bar{Q}$ mass \cite{manohar}. Unlike the pole mass, the $1S$ mass, containing,  
by construction,  half of the total static energy $\langle 2m +V^{\rm Coul}\rangle$,  
is free of ambiguities of order $\Lambda_{\rm QCD}$. Our strategy will be the following. 
First, we consider the perturbative contribution (up to order $\alpha_s^4$) 
of the $\, ^3S_1$ levels of charmonium and bottomonium:
$$
E(J/\psi)_{\rm  pert}= f(m_c), \quad \quad \quad  E(\Upsilon(1S))_{\rm  pert}=f(m_b), 
$$
which are respectively a function of the $c$ and the $b$ pole mass and can be read off from 
Eq. (\ref{EBc2}) in the equal-mass case, adding to it the spin-spin interaction energy: 
$m(C_F\alpha_{\rm s})^4/3$. We invert these relations in order to obtain the pole masses 
as a formal perturbative expansion depending on the $1S$ mass. Finally, we insert the expressions 
$m_c=f^{-1}(E(J/\psi)_{\rm  pert})$ and $m_b=f^{-1}(E(\Upsilon(1S))_{\rm  pert})$ in  Eq. (\ref{EBc2}). 
At this point we have the perturbative mass of the $B_c$ as a function of the  
$J/\psi$ and $\Upsilon(1S)$ perturbative masses
\begin{equation}
E(B_c)_{\rm  pert}= f(f^{-1}(E(J/\psi)_{\rm  pert}),f^{-1}(E(\Upsilon(1S))_{\rm  pert})).
\label{bc}
\end{equation}
If we identify the perturbative masses $E(J/\psi)_{\rm  pert}$, $E(\Upsilon(1S))_{\rm  pert}$ with the physical 
ones, i.e. $E(J/\psi)_{\rm  phys} = 3097$ MeV and $E(\Upsilon(1S))_{\rm  phys} = 9460$ MeV \cite{pdg}, 
then the expansion (\ref{bc}) depends only on the scale $\mu$. \vspace{2mm}

\begin{figure}[htb]
\makebox[-1truecm]{\phantom b}
\put(70,0){\epsfxsize=10truecm\epsffile{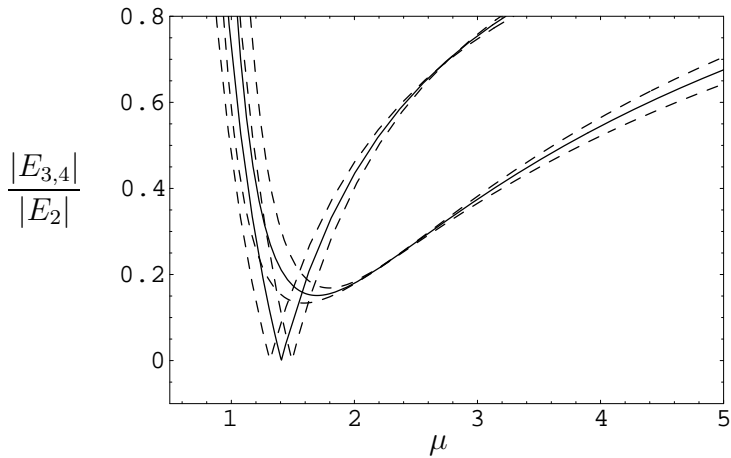}}
\put(80,100){$\displaystyle{|E_{3,4}|\over |E_2|}$}
\put(240,5){$\mu$}
\vskip 0.2truecm
\caption{$|E_3/E_2|$ and $|E_4/E_2|$ as a function of $\mu$, being $E_n$  the
  order $\als^n$ contribution to $E(B_c)_{\rm  pert}$. The continuous lines 
  correspond to $\Lambda_{\overline{\rm MS}}^{N_f=3}=300$ MeV, the dashed 
  lines to $\Lambda_{\overline{\rm MS}}^{N_f=3}=250$ and
  $\Lambda_{\overline{\rm MS}}^{N_f=3}=350$ respectively. $E_3$ vanishes 
  for values of $\mu$ around 1.4 GeV.} 
\vskip 0.6truecm
\label{plotper}
\end{figure}

In Fig. \ref{plotper} we show the dependence on $\mu$ of the order $\als^3$ and 
$\als^4$ contributions to $E(B_c)_{\rm  pert}$ respectively.
Taking into account that the order $\als^3$ contribution vanishes at $\mu
\simeq 1.4$ GeV, the perturbative series seems to be reliable  
for values of $\mu$ bigger than $(1.2 \div 1.3)$ GeV and 
lower than $(2.6 \div 2.8)$ GeV. For instance, $E(B_c)_{\rm  pert} = 6278.5 +  35 + 6.5 + 5.5$ MeV 
at the scale $\mu=1.6$ GeV. This is consistent with: {\it i)} the fact that, 
for values of $\mu$ close to or less than 1 GeV, the perturbative calculation 
(and the initial assumption that $B_c$ is Coulombic or quasi-Coulombic)
is expected to break down; {\it ii)} the fact that higher values of $\mu$ do not correspond
to the characteristic scale of the system (this is signalled by the appearance
of big logarithms in the perturbative expansion); {\it iii)} the estimate of
the scale $\mu$ inferred in the introduction from the size of the $J/\psi$. 
More precisely we will take in our analysis\footnote{
The inclusion of a somewhat higher energy region, which seems to be allowed by 
Fig. \ref{plotper}, would not change our final result (\ref{ebc}). E.g. taking  
1.2 GeV $\le \mu \le$  2.6 GeV we would get, by keeping the same central values 
as above, $E(B_c)_{\rm  pert} = 6326^{+29}_{-10}\, {\rm MeV}$.}
1.2 GeV $\le \mu \le$  2.0 GeV and 250 MeV $\le \Lambda_{\overline{\rm MS}}^{N_f=3} \le$ 
350 MeV (in terms of $\als$ this corresponds to $0.26 \siml \, \als(2\,{\rm GeV}) \siml \, 0.30$). 
In this way we entirely cover the energy range used in
\cite{Pineda} in order to study the $J/\psi$.\footnote{
Actually the range considered in \cite{Pineda} was 1.36 GeV $\le \mu \le$  1.76 GeV.}

\begin{figure}[htb]
\makebox[-1truecm]{\phantom b}
\put(70,0){\epsfxsize=10truecm\epsffile{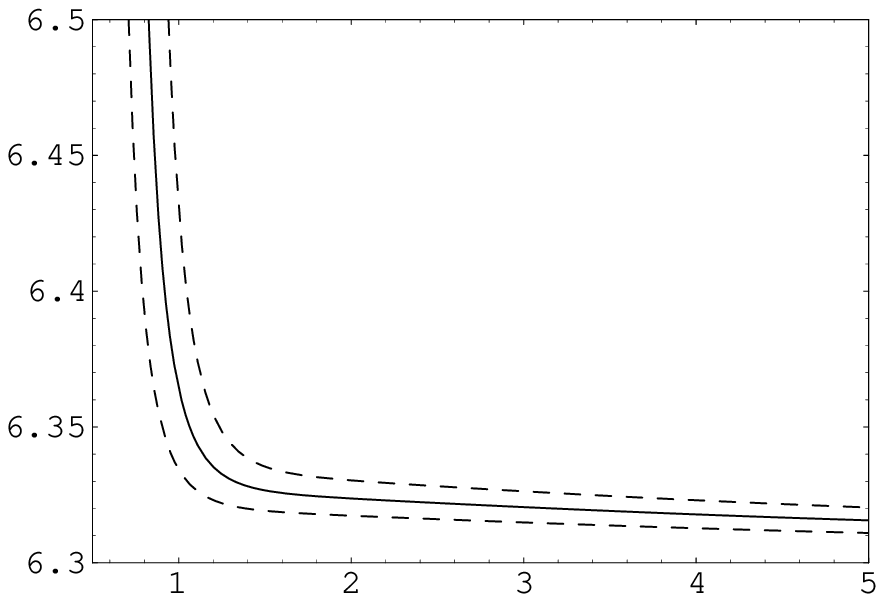}}
\put(55,100){$E(B_c)_{\rm pert}$}
\put(240,5){$\mu$}
\vskip 0.2truecm
\caption{$E(B_c)_{\rm pert}$ as a function of $\mu$ at $\Lambda_{\overline{\rm
      MS}}^{N_f=3}=300$ MeV (continuous line).  The dashed lines refer 
      to $\Lambda_{\overline{\rm MS}}^{N_f=3}=250$ MeV and
      $\Lambda_{\overline{\rm MS}}^{N_f=3}=350$ MeV respectively.}
\vskip 0.6truecm
\label{plot}
\end{figure}

By varying $\mu$ from 1.2 GeV to 2.0 GeV and $\Lambda_{\overline{\rm MS}}^{N_f=3}$ 
from 250 MeV to 350 MeV and by calculating the maximum variation 
of $E(B_c)_{\rm  pert}$ in the given range of parameters, we get 
as our final result  
\begin{equation}
E(B_c)_{\rm  pert} = 6326^{+29}_{-9}\, {\rm MeV}. 
\label{ebc}
\end{equation}
The upper limit corresponds to the choice of parameters  $\Lambda_{\overline{\rm
MS}}^{N_f=3}=350$ MeV, $\mu=1.2$ GeV, while the lower limit to  
$\Lambda_{\overline{\rm MS}}^{N_f=3}=250$ MeV and $\mu = 2.0$ GeV. 
As a consequence of the now obtained good behaviour of the perturbative series 
in the considered range of parameters, our result appears stable with respect 
to variations of $\mu$ (see Fig. \ref{plot}).
We would like to note that the main source of error in Eq. (\ref{ebc}) 
comes from the border region 1.2 GeV $\siml \, \mu <$  1.4 GeV at 
$\Lambda_{\overline{\rm MS}}^{N_f=3} \siml$ 350 MeV, where it may become questionable to treat 
the $B_c$ as a Coulombic or quasi Coulombic system (see Fig. \ref{plotper}).

\section{Discussion and Conclusions}
We have calculated the perturbative $B_c$ mass as defined by Eq. (\ref{EBc}). 
The problem of the bad behaviour of the perturbative 
series has been overcome by expressing the perturbative $B_c$ mass in terms of the perturbative 
$J/\psi$ and $\Upsilon(1S)$ masses. The series we obtain has a good
convergent behaviour. This fact is relevant since it shows that the scale
hierarchy considered in the introduction ($m > m v > \lQ$), which led to
the Hamiltonian (\ref{bc2}), correctly applies to the system under
consideration.\footnote{For instance, an analogous analysis carried out on the $B_s$ 
system does not show any sign of convergence.}
In other words, the result we get is consistent with the
assumption made that the $B_c$ system is Coulombic or quasi-Coulombic. 
Moreover, the perturbative series turns out to 
be weakly sensitive to variations of $\mu$ (the renormalization scale) and 
$\Lambda_{\overline{\rm MS}}^{N_f=3}$ in the range 1.2 GeV $\le \mu \le$ 2.0 GeV 
and 250 MeV $\le \Lambda_{\overline{\rm MS}}^{N_f=3} \le$ 350 MeV. 
The result appears, therefore, reliable from a perturbative 
point of view. Non-perturbative contributions have not been taken into account so far. 
They affect the identification of the perturbative masses 
$E(B_c)_{\rm  pert}$, $E(\Upsilon(1S))_{\rm  pert}$, 
$E(J/\psi)_{\rm  pert}$, with the corresponding physical ones through Eq. (\ref{EBc2}). 
Let us call these non-perturbative contributions $\delta E(B_c)$, 
$\delta E(\Upsilon(1S))$ and $\delta E(J/\psi)$ respectively. 
As discussed in the introduction, depending on the actual kinematic situation of the system, 
they can be of potential or non-potential nature. 
In the last case they can be encoded into non-local condensates or into local condensates.  
There is no way to discriminate among these situations, since the size of what
would be the energy scale $m v^2$ of the system with respect to $\lQ$ is unknown. 
Non-perturbative contributions affect the identification of Eq. (\ref{bc}) with the 
physical $B_c$ mass roughly by an amount $\simeq - \displaystyle {\delta E(J/\psi)\over 2}$ 
$- \displaystyle {\delta E(\Upsilon(1S))\over 2}$  $+ \delta E(B_c)$. 
Assuming $|\delta E(J/\psi)| \le 300$ MeV, $|\delta E(\Upsilon(1S))| \le 100$ MeV 
and $\delta E(\Upsilon(1S)) \le \delta E(B_c) \le \delta E(J/\psi)$, the 
identification of our result (\ref{ebc}) with the physical $B_c$ mass may, in principle, 
be affected by uncertainties, due to the unknown non-perturbative 
contributions, as big as $\pm 200$ MeV. However, the different $\delta E$ are correlated, 
so that we expect, indeed, smaller uncertainties. If we assume, for instance, 
$\delta E(\Upsilon(1S))$ and $\delta E(J/\psi)$ to have the same sign, which seems to be 
quite reasonable, then the above uncertainty reduces to $\pm 100$ MeV. 
Constraining even more the form of $\delta E$, by evaluating it from the Voloshin--Leutwyler 
formula (i.e. in terms of local condensates), as given in Ref. \cite{Pineda}, we get a negative 
contribution (since the term coming from the $J/\psi$ is the dominating one) of less than $100$ MeV. 
This feature, if preserved also in the other kinematic situations, would confirm, indeed, that 
the effect of the non-perturbative contributions is not too big and that its effect is to lower 
down the perturbative result given in Eq. (\ref{ebc}).
More quantitative statements are difficult to make, since, differently from 
the perturbative case discussed in the previous section,  they appear to be 
dependent on the choice of the parameters. 

The result we ~~get in ~~Eq. (\ref{ebc}) ~~ is compatible with the experimental value 
$E(B_c)_{\rm phys}$ $=$ $6.40 \pm 0.39 \pm 0.13$ GeV reported in \cite{exp}.
We mention that OPAL reports in \cite{opal} 2 candidates $B_c$ in hadronic 
$Z^0$ decays events, with an estimated mass $E(B_c)_{\rm phys} = 6.32 \pm 0.06$ GeV.
Also this value compares favourably with ours. Having more precise and established experimental data 
will make it possible to make some more definite statements. In particular, it will be possible to 
give, inside a Coulombic or quasi-Coulombic picture, a precise estimate of the size of the non-perturbative 
effects in the $B_c$ mass. In the table, we also report, for comparison, some of the other 
determinations of the $B_c$ mass available in the literature. The results 
quoted in \cite{quigg,kiselev,fulcher,sumrule} rely on potential models 
(essentially a Coulomb plus a confining potential) and  
are reported without errors. The figure that appears in the table in correspondence of 
Ref. \cite{sumrule} refers to an average of different models performed 
by those authors. Finally \cite{lattice} reports the result of a very recent 
lattice calculation.  We would like to note that, if one assumes that potential models 
give a $B_c$ mass close to reality, then, comparing the potential model
predictions with Eq. (\ref{ebc}), non-perturbative contributions seem to be of 
the order $- (40 \div 100)$ MeV (consistently with expectations,
non-perturbative corrections become as smaller as perturbation theory better works,
i.e. in correspondence of low values of $\Lambda_{\overline{\rm MS}}^{N_f=3}$ 
and high values of $\mu$). Finally, it is interesting to notice that these 
figures are completely consistent with the general discussion on the uncertainties,  
coming from non-perturbative contributions, done above.

\begin{table}[h]
\vspace{-2mm}
\makebox[2cm]{\phantom b}
\begin{tabular}{|l|l|}
\hline
\hbox{\hspace{0.8cm}}Obtained from \hbox{\hspace{4cm}}& $B_c(1^{1}\!S_{0})$ in MeV \hbox{\hspace{5cm}}\\
\hline
\hbox{\hspace{0.8cm}}Experiment \cite{exp}             & $6400\pm 390 \pm 130$ \\
\hbox{\hspace{0.8cm}}Eq. (\ref{ebc})   (pert. mass)    & $6326^{+29}_{-9}$      \\
\hbox{\hspace{0.8cm}}\cite{quigg}                      & 6264 \\
\hbox{\hspace{0.8cm}}\cite{kiselev}                    & 6253 \\
\hbox{\hspace{0.8cm}}\cite{fulcher}                    & 6286 \\
\hbox{\hspace{0.8cm}}\cite{sumrule}                    & 6255 \\
\hbox{\hspace{0.8cm}}\cite{lattice}                    & 6.386(9)(98)(15) \\
\hline
\end{tabular}
\vspace{2mm}
\label{tab1}
\caption{Mass of the $B_{c}$ meson. 
The result labelled with Eq. (\ref{ebc}) refers to the present work.}
\end{table}

\section*{Acknowledgments}
The authors thank Andre Hoang and Antonio Pineda for valuable comments and suggestions.


\begin{references}
\bibitem{exp}  F. Abe et al., CDF Collaboration, Phys. Rev. Lett. {\bf 81}, 2432 (1998).
\bibitem{pnrqcd2} N. Brambilla, A. Pineda, J. Soto and A. Vairo,
  Nucl. Phys. {\bf B 566}, 275 (2000).   
\bibitem{pnrqcd1} N. Brambilla, A. Pineda, J. Soto and A. Vairo, Phys. Lett. {\bf B 470}, 215 (1999).   
\bibitem{Voloshin} M. B. Voloshin, Nucl. Phys. {\bf B 154}, 365 (1979); 
 H. Leutwyler, Phys. Lett. {\bf B 98}, 447 (1981).
\bibitem{Pineda} S. Titard and F. J. Yndurain, Phys. Rev. {\bf D 49}, 6007 (1994); 
 A. Pineda and F. J. Yndurain, Phys. Rev. {\bf D 58}, 094022 (1998);
 {\bf D 61}, 077505 (2000).
\bibitem{Pineda2} A. Pineda, Nucl. Phys. {\bf B 494}, 213 (1997).
\bibitem{pot} N. Brambilla and A. Vairo, hep-ph/9904330; G. Bali, hep-ph/0001312.
\bibitem{wilson} K. G. Wilson, Phys. Rev. {\bf D 10}, 2445 (1974);
 E. Eichten and  F. L. Feinberg, Phys. Rev. Lett. {\bf 43}, 1205 (1979);
 D. Gromes, Z. Phys. {\bf C 26}, 401 (1984); 
 A. Barchielli, E. Montaldi and G. M. Prosperi, Nucl. Phys. {\bf B 296}, 625 (1988) 
 and Erratum ibid. {\bf B 303}, 752 (1988); 
 A. Barchielli, N. Brambilla and G. Prosperi, Nuovo Cimento {\bf 103 A}, 59 (1990);
 N. Brambilla, A. Pineda, J. Soto and A. Vairo, hep-ph/0002250.
\bibitem{Brambilla} N. Brambilla and A. Vairo, Phys. Rev. {\bf D 55}, 3974 (1997).
\bibitem{beneke}  M. Beneke, Phys. Lett. {\bf B 434}, 115 (1998); 
 A. Hoang, M. Smith, T. Stelzer and S. Willenbrock, Phys. Rev. {\bf D 59}, 114014 (1999); 
 A. Pineda, Ph. D. Thesis, (Barcelona), Jan. 1998; N. Uraltsev, hep-ph/9804275. 
\bibitem{manohar} A. H. Hoang, Z. Ligeti and A. V. Manohar, Phys. Rev. Lett. {\bf 82}, 277 (1999);
 A. Hoang, hep-ph/9909356.
\bibitem{cloop} A. H. Hoang and A. V. Manohar, Phys. Lett. {\bf B 483}, 94 (2000).
\bibitem{Peter}  Y. Schr\"oder, Phys. Lett. {\bf B 447}, 321 (1999);
 M. Peter, Phys. Rev. Lett. {\bf 78}, 602 (1997).
\bibitem{Gupta} S. N. Gupta and S. F. Radford, Phys. Rev. {\bf D 25}, 3430 (1982).
\bibitem{pdg} C. Caso et al.,  Particle Data Group, Eur. Phys. J. {\bf C 3}, 1 (1998); http://pdg.lbl.gov/. 
\bibitem{quigg} E. Eichten and C. Quigg, Phys. Rev. {\bf D 49}, 5845 (1994).
\bibitem{kiselev} S. S. Gershtein, V. V. Kiselev, A. K. Likhoded and A. V. Tkabladze, Phys. Usp. 38, 1 (1995); 
 a calculation which leads to the same numerical value for the $B_c$ mass can be found in 
 V. O. Galkin, A. Yu. Mishurov and R. N. Faustov, Sov. J. Nucl. Phys. {\bf 55}(8), 1207 (1992).  
\bibitem{fulcher} L. P. Fulcher, Phys. Rev. {\bf D 60}, 074006 (1999).
\bibitem{sumrule} E. Bagan, H. G. Dosch, P. Gosdzinsky, S. Narison and J. M. Richard, 
 Z. Phys. {\bf C 64}, 57 (1994).  
\bibitem{lattice} H. P. Shanahan, P. Boyle, C. T. H. Davies and H. Newton, Phys. Lett. {\bf B 453}, 289 (1999).
\bibitem{opal} K. Ackerstaff et al., Phys. Lett. {\bf B 420}, 157 (1998).
\end{references}
\end{document}